\documentclass[%
 reprint,
superscriptaddress,
 amsmath,amssymb,
 aps,prl
]{revtex4-1}
\usepackage{graphicx}

\begin{document}
\title{Infrasonic wave propagation in ultrasoft solids at low Reynolds numbers }

\author{Jan Maarten van Doorn}
\affiliation{%
 Physical Chemistry and Soft Matter, Wageningen University, Stippeneng 4, 6708 WE, Wageningen, The Netherlands
}
\author{Ruben Higler}
\affiliation{%
 Physical Chemistry and Soft Matter, Wageningen University, Stippeneng 4, 6708 WE, Wageningen, The Netherlands
}
\author{Ronald Wegh}
\affiliation{%
 Physical Chemistry and Soft Matter, Wageningen University, Stippeneng 4, 6708 WE, Wageningen, The Netherlands
}
\author{Remco Fokkink}
\affiliation{%
 Physical Chemistry and Soft Matter, Wageningen University, Stippeneng 4, 6708 WE, Wageningen, The Netherlands
}
\author{Alessio Zaccone}
\affiliation{%
 Department of Chemical Engineering and Biotechnology, University of Cambridge, Cambridge CB3 0AS, UK
}
\author{Joris Sprakel}
\affiliation{%
 Physical Chemistry and Soft Matter, Wageningen University, Stippeneng 4, 6708 WE, Wageningen, The Netherlands
}
\author{Jasper van der Gucht}
\email{jasper.vandergucht@wur.nl}
\affiliation{%
 Physical Chemistry and Soft Matter, Wageningen University, Stippeneng 4, 6708 WE, Wageningen, The Netherlands
}

\date{\today}

\begin{abstract}
The propagation of elastic waves in soft materials plays a crucial role in the spatio-temporal transmission of mechanical signals, e.g. in biological mechanotransduction\cite{Ingber2010,jaalouk2009mechanotransduction} or in the failure of marginal solids\cite{ PhysRevLett.95.098301, PhysRevLett.108.058001, broedersz2011criticality,during2013phonon}.  At high Reynolds numbers $Re \gg 1$, inertia dominates and wave propagation can be readily observed\cite{Buttinoni201712266,Grasland-Mongrain861,mallet1846dynamics}. However, mechanical cues in soft and biological materials often occur at low $Re$ \cite{doi:10.1146/annurev-conmatphys-070909-104120}, where waves are overdamped. Not only have low $Re$ waves been difficult to observe in experiments, their theoretical description remains incomplete. In this paper, we present direct measurements of low $Re$ waves propagating in ordered and disordered soft solids, generated by an oscillating point force induced by an optical trap. We derive an analytical theory for low $Re$ wave propagation, which is in excellent agreement with the experiments. Our results present both a new method to characterize wave propagation in soft solids and a theoretical framework to understand how localized mechanical signals can provoke a remote and delayed response.
\end{abstract}

\maketitle
The spatiotemporal response to mechanical perturbations is one of the key factors that determines the fate of soft materials\cite{PhysRevLett.103.036001, goyon2008spatial}. For example, in marginally-stable systems, such as jammed packings or fiber networks, a stress at the right position can cause total loss of rigidity\cite{ PhysRevLett.95.098301, PhysRevLett.108.058001, broedersz2011criticality,during2013phonon}. Also in living cells, the propagation of mechanical signals through soft structures is crucial in mechanotransduction\cite{Ingber2010,jaalouk2009mechanotransduction}, and controls, for example, cell dif-ferentiation\cite{Discher1139}. Elastic wave propagation is a prototypic example of how a localised mechanical signal can spread in space and time, and is intimately linked to the mechanical properties of the medium\cite{auld1973acoustic}. However, it is governed by more complex mechanisms than the simple sum of elastic and viscous responses \cite{Rogers2018}. The complex viscoelastic response of soft materials becomes particularly apparent upon mechanical excitation at Deborah numbers $De = \omega \tau \approx 1$, where $\omega$ is the excitation frequency  and $\tau$ the intrinsic relaxation time  of the solid\cite{poole2012deborah}. Excitation at these low frequencies also implies low Reynolds numbers where viscous attenuation of the wave signal is strong and their detection challenging. Moreover, in ultrasoft solids the relative amplitudes of thermal fluctuations are large, thus further obscuring accurate wave detection in experiments. 
In this Letter we show how Fourier filtering can reveal even very weak propagating elastic waves at extremely low Reynolds numbers, $Re \sim 10^{-6}$, in ultrasoft solids, formed from crystals and glasses of colloids in two dimensions. We create a localized oscillatory perturbation within these solids with an optical tweezer and use video microscopy and frequency-domain filtering to quantify the spatiotemporal strain response. On the basis of an overdamped equation of motion, which is in excellent quantitative agreement with our experimental results, this enables a full characterization of the linear mechanics of even very weak elastic solids. 

\begin{figure}[t!]
    \centering
    \includegraphics[width=\linewidth]{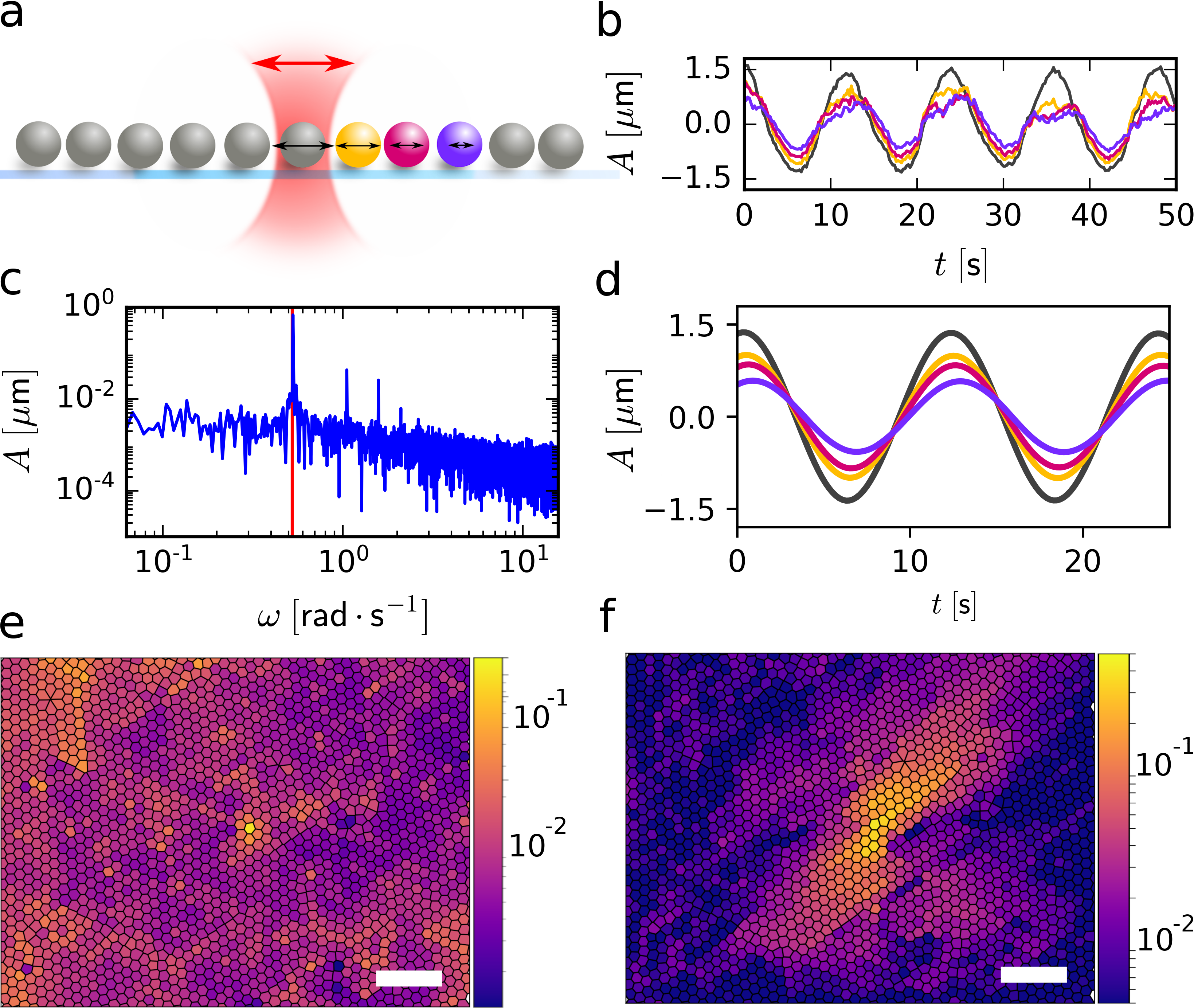}
    \caption{(a) Schematic overview of our experiment. (b) Raw experimental particle trajectories in the direction of the driven particle (grey), colors correspond to the particles in (a). (c) Amplitude spectrum of the driven particle, the red line corresponds to the driving frequency. (d) Trajectories of (b) after line filtering. (e) Unfiltered root-mean-square displacement and (f) Fourier-filtered amplitude map of particles in a crystal excited at 3.1 rad/s; the color scale in (e) and (f) represents the amplitude in $\mu$m, scale bars represent 40$\mu$m.}
    \label{fig:1}
\end{figure}

We prepare two-dimensional hexagonal crystals by sedimenting mono-disperse silica particles with diameter $d=6.25\,\mu$m suspended in an aqueous solution. This yields dense crystals with long-ranged hexagonal order at a packing fraction of 0.89 (Supplementary Information 2).

To create a propagating mechanical wave, we trap a single particle of the crystal in an optical trap and force it into an in-plane oscillatory motion with an amplitude of 2.5$\,\mu$m (Fig.\ref{fig:1}a). We vary the frequency of this motion between 0.05 and 10 rad/s, corresponding to $2 < De < 2\cdot10^2$ and $Re\approx3\cdot$10$^{-6}$ and confirm that this mechanical excitation is well within the linear regime (Supplementary Information 1, 5). 

\begin{figure}[t!]
    \centering
    \includegraphics[width=\linewidth]{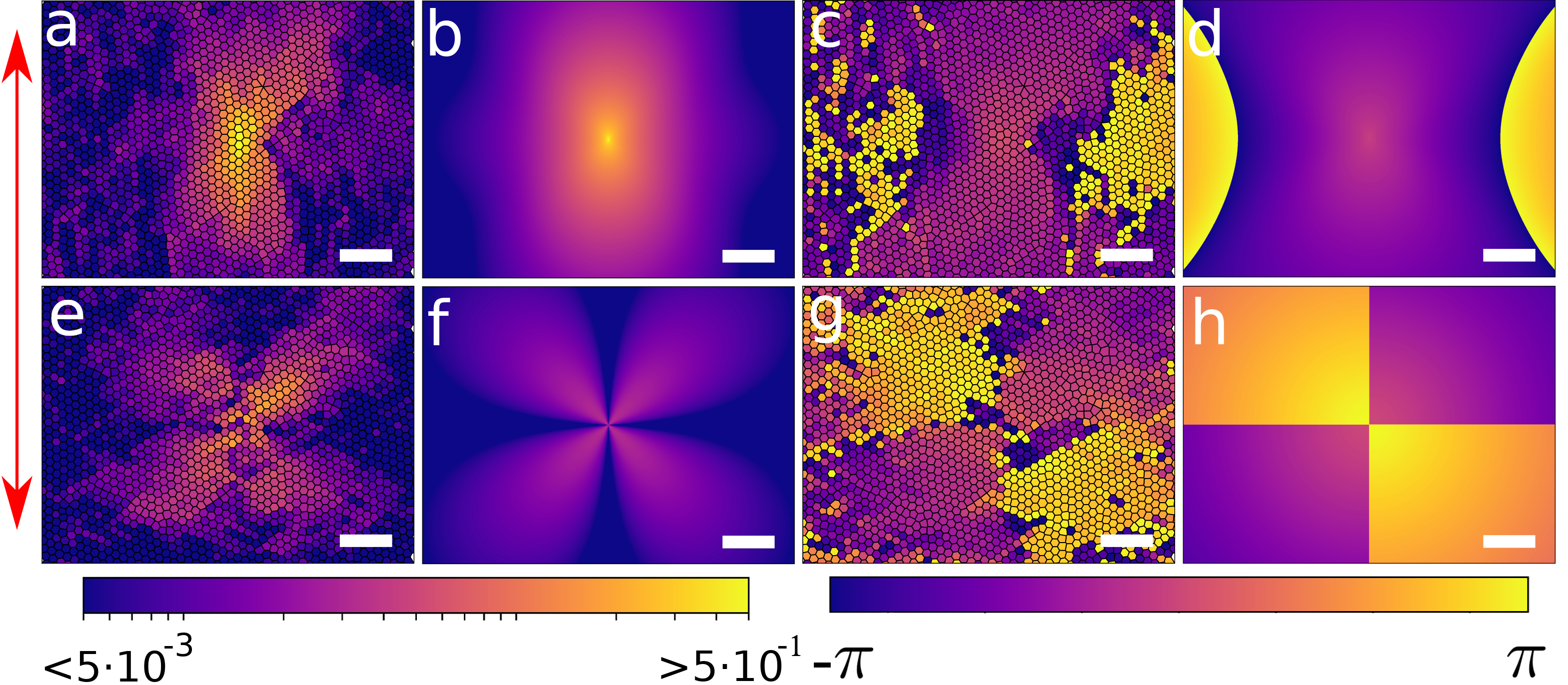}
    \caption{(a) Experimental amplitude map of the parallel displacement. (b) Parallel displacement amplitude predicted by our model. (c) Experimental phase shift for the parallel displacement. (d) Phase shift of parallel displacement predicted by our model.(e) Experimental map of perpendicular displacement. (f) Perpendicular displacement amplitude predicted by our model. (g) Experimental phase shift for the perpendicular displacement.(h) Phase shift of perpendicular displacement predicted by our model. Amplitudes and phases have units micron and radians respectively, scalebars represent 40$\mu$m, the red arrow indicates the oscillation direction.}
    \label{fig:2}
\end{figure} 

The oscillating particle creates a mechanical wave that propagates through the surrounding material. This sets up a net ballistic displacement of the particles adjacent to the oscillating bead. However, the signal of interest is convoluted with the inherent Brownian motion of these microscopic colloids. Especially far away from the trapped colloid, where the elastic signal is attenuated, it may drown in the Brownian noise (Fig.\ref{fig:1}b,e). Upon filtering the positional trajectory of each particle in the frequency domain at the driving frequency (Fig.\ref{fig:1}c), even small displacements due to the propagating wave become apparent (Fig.\ref{fig:1} d). To ensure statistical reliability of these data, we set-up a real-time distributed particle tracking algorithm that allows us to collect data during 20 000 -- 35 000 frames, which is equivalent to 50--500 oscillation cycles. While the unfiltered mean-square displacement of the particles shows no apparent signature of the perturbation, except at the forced particle (Fig.\ref{fig:1} e), the Fourier-filtered amplitude map clearly shows a propagating mechanical wave with an amplitude that decays steeply with increasing distance from the trapped bead (Fig.\ref{fig:1} f) and a phase shift that gradually increases with distance (Fig.\ref{fig:1} d).
 
To explain these results we assume that the colloidal crystal can be treated as a two-dimensional continuous elastic material. We write an equation of motion for the displacement field $\vec{u}$ in the solid \cite{toell} to which we add a dissipative term to account for the damping fluid and an oscillating point force $\vec{f}(t) = \vec{f}_{0} \cdot e^{i \omega t}$ that represents the perturbation:
\begin{equation}
\rho \frac{\partial^2 \vec{u} }{\partial t^2} + \gamma  \frac{\partial \vec{u}}{\partial t}= \vec{f}(t) + \frac{E}{2(1+\nu)} \vec{\nabla}^2\vec{u} + \frac{E}{2(1-\nu)}\vec{\nabla}(\vec{\nabla} \cdot \ \vec{u})
\label{EquationOfMotion}
\end{equation}
Here the first term describes the inertial forces with $\rho$ the density of the 2D material. The second term represents the viscous damping due to the solvent with $\gamma$ the drag coefficient per unit area, which we determine experimentally to be $\gamma=2.9\cdot10^3$ Ns/m$^3$ (Supplementary Information 5). The last two terms correspond to the Navier-Cauchy equation that describes the elastic forces within the 2D solid with $E$ the 2D elastic modulus and $\nu$ the Poisson ratio.

Since our experiments are performed at low Reynolds number, the inertial term is negligible, resulting in overdamped mechanics. Solving the equation of motion for this case yields the displacement field in the form $\vec{u} = \bm{\alpha}\cdot\vec{f}$
with  $\bm{\alpha}$ the complex response function, which has components $\alpha_{\parallel}$ and $\alpha_{\bot}$ that describe the components of the displacement field parallel and perpendicular to the applied force, respectively. In polar coordinates, with $r$ the distance from the point where the force is applied and $\theta=0$ corresponding to the direction of the force, this becomes (see Supplementary Information 7 for full details):
\begin{multline}\label{eq:alphapar}
    \alpha_{\parallel}(r,\theta) = \frac{1-\nu^2}{4\pi E}\left(K_{0}\Big(\frac{r\sqrt{i}}{\zeta}\Big)+\lambda^2K_{0}\Big(\frac{r\sqrt{i}}{\lambda\zeta}\Big)+\right.\\
    \left.\cos{(2\theta)}\left[K_{2}\Big(\frac{r\sqrt{i}}{\zeta}\Big)-\lambda^2K_2\Big(\frac{r\sqrt{i}}{\lambda\zeta}\Big)\right]\right)
\end{multline}
and
\begin{equation}\label{eq:alphaperp}
    \alpha_{\perp}(r,\theta) = \frac{1-\nu^2}{4\pi E}\sin(2\theta)\left[K_2\Big(\frac{r\sqrt{i}}{\zeta}\Big))-\lambda^2K_2\Big(\frac{r\sqrt{i}}{\lambda\zeta}\Big)\right]
\end{equation}
where $K_0$ and $K_2$ denote modified Bessel functions of the second kind,  $\zeta=(\omega\gamma(1-\nu^2)/E)^{-\frac{1}{2}}$ is a characteristic attenuation length of the displacement amplitude, and $\lambda=\sqrt{2/(1-\nu)}$ is a parameter that depends only on the Poisson ratio and diverges for $\nu \rightarrow{} 1$ which is the maximum Poisson ratio in 2D.  The amplitude and the phase of the displacement fields are obtained as the magnitude and the argument, respectively, of the complex response functions.

To compare our experimental results with this prediction, we decompose the measured displacement amplitudes into  their parallel and perpendicular components (Fig. \ref{fig:2}a and e). This results in distinct lobed patterns for these two components that can be observed in all of our experiments. Our theoretical prediction produces identical patterns that are in excellent agreement, indicating that wave propagation in these colloidal crystals can indeed be described by treating the material as a continuous 2D elastic solid (Fig. \ref{fig:2}b and f). 

The parallel component of the displacement response propagates preferentially along the excitation axis and shows a distinct asymmetry in the attenuation length along the two primary axes. 
The perpendicular displacement shows a four-lobed pattern, with maximum displacements at an angle of 45$^\circ$ with respect to the excitation direction. 
Also in the phase maps (Fig.\ref{fig:2}c,d), we observe patterns that are in excellent agreement with the theoretical prediction (Fig.\ref{fig:2}g,h). 

\begin{figure}[t!]
    \centering
    \includegraphics[width=\linewidth]{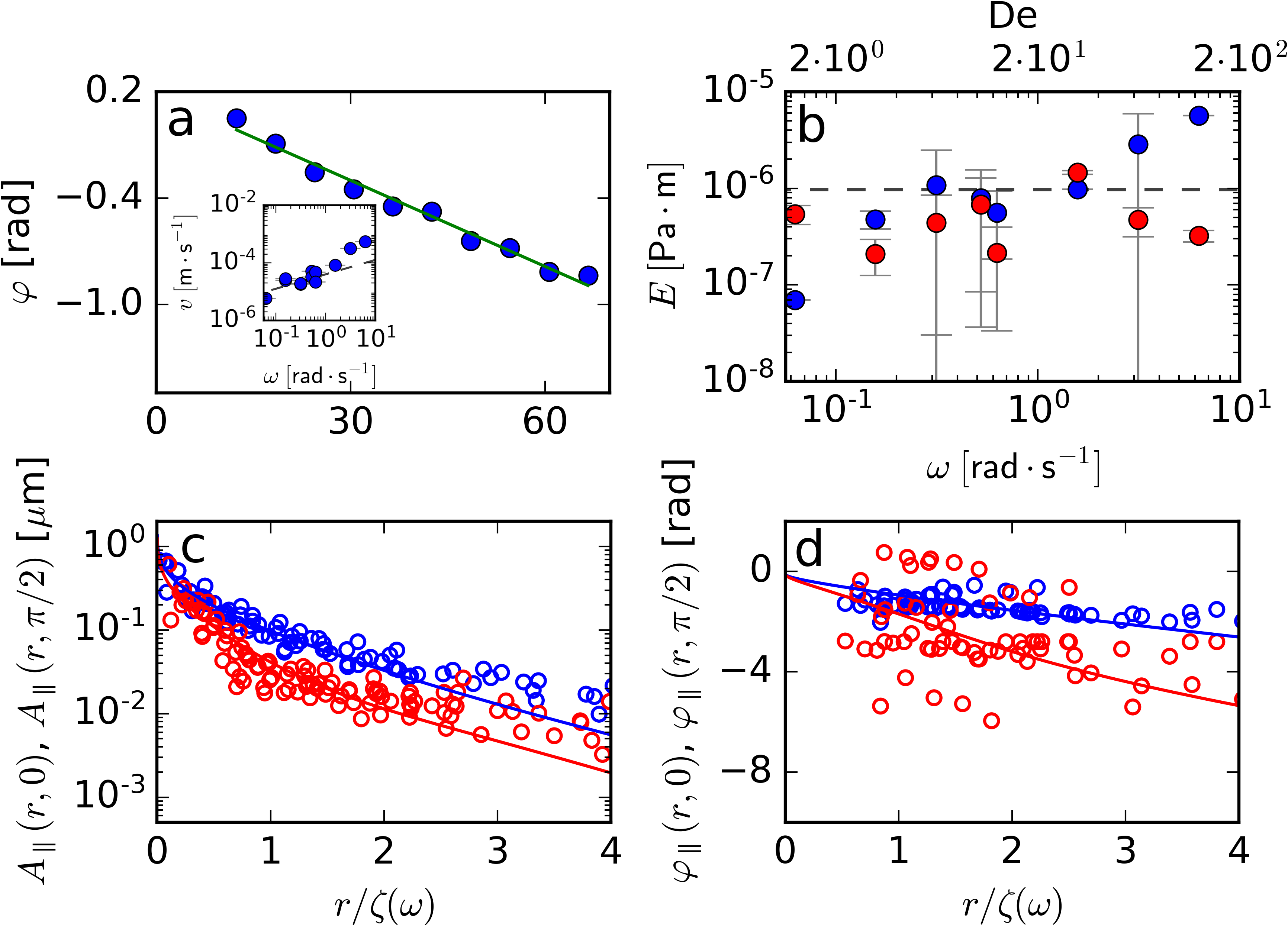}
    \caption{(a) Bin-averaged phase of the parallel displacement in the direction of the excitation for $\omega = 0.52$ rad/s . Inset shows the phase velocity versus probed frequency; dashed line depicts slope $\frac{1}{2}$, errorbars depict a 95\% confidence interval. (b) Elastic modulus as a function of frequency, obtained from the phase (blue) and amplitude (blue) of the parallel displacement components; errorbars depict a 95\% confidence interval and dashed line indicates affine prediction. (c) Superposition of parallel displacement amplitude for different frequencies along $\theta=0$ (blue) and $\theta=\pi/2$ (red) versus normalized distance. (d) Superposition of parallel displacement phase for different frequencies along $\theta=0$ (blue) and $\theta=\pi/2$ (red) versus normalized distance. Lines in c and d represent the theoretical predictions.}
    \label{fig:3}
\end{figure}

We can now use our theoretical analysis to interpret the experimental results in terms of the linear elasticity of the solid. For this, we consider the parallel displacement component in the direction of the excitation, $\theta=0$. An asymptotic expansion of Equation \ref{eq:alphapar} for relatively large distances from the perturbation $r>\zeta$ (see Supplementary Information 7) leads to a phase lag in the far field
\begin{equation}\label{eq:phasepar}
    \phi_\parallel(r,0)\approx -\frac{\pi}{8}-\frac{r}{\zeta\sqrt{2}}
\end{equation}
and an amplitude 
\begin{equation}\label{eq:amplitudepar}
   A_\parallel(r,0)\sim r^{-1/2}e^{-r/\zeta\sqrt{2}}
\end{equation}
According to Equation \ref{eq:phasepar}, the phase varies linearly with $r$ along $\theta=0$, which is indeed what we find experimentally (Fig. \ref{fig:3}a). This means that the wave propagates at a constant velocity in the direction of the excitation, with a phase velocity $v_\parallel=|\omega(d\phi_\parallel/dr)^{-1}|\approx\omega\zeta\sqrt{2}=\sqrt{2E\omega/\gamma(1-\nu)^2}$. As shown in the inset of Fig. \ref{fig:3}a, the phase velocity increases approximately as $v_\parallel\sim\sqrt{\omega}$ for low frequencies, which indicates that the elastic modulus and the Poisson ratio do not depend on the frequency in this regime. Using Equations \ref{eq:phasepar} and \ref{eq:amplitudepar} we can determine the characteristic length $\zeta$ as a function of frequency in two independent ways. Taking a value of the Poisson ratio $\nu=0.5$ (see below), we can then estimate  the elastic modulus of the colloidal crystal. Both the phase and amplitude data give moduli in the range $E\approx10^{-6}$ N/m (Fig. \ref{fig:3}b). These values for $E$ may be compared with a simple estimate obtained by  approximating the colloidal crystal as a hexagonal lattice of harmonic springs, for which $E=2k/\sqrt{3}$ with $k$ the spring constant of a particle pair \cite{boal_2012}. We estimate $k$ by analyzing the thermal bond length fluctuations (Supplementary Information 4) and find $E\approx1\cdot10^{-6}$ N/m in very good agreement with the experimental values (dashed line in Fig. \ref{fig:3}b). 

We note that our method is limited at higher frequencies by the decrease in the attenuation length with increasing frequency. Once the characteristic length $\zeta$ becomes of the order of the particle size, discretization effects hinder the accurate determination of the wave propagation. This gives a limiting frequency $\omega_{max}\approx E/\gamma d^2\approx10$ rad/s.

The linear elasticity of a two-dimensional solid is described by two independent mechanical parameters, the elastic modulus  and the Poisson ratio. While the modulus only affects the absolute magnitude of the response function and the characteristic length scale $\zeta$, the shape of the spatial pattern is uniquely determined by the Poisson ratio, as expressed by the parameter $\lambda$ in Equation \ref{eq:alphapar} (SI Fig. 26). It should therefore be possible to superimpose the measured displacement data obtained at different frequencies by normalizing the distance $r$ by the characteristic length $\zeta$. Figure \ref{fig:3}c and d show such a collapse for the amplitude and the phase, respectively, of the parallel displacement components in the direction of the excitation ($\theta=0$) and perpendicular to it ($\theta=\pi/2$). We fit these curves to equation \ref{eq:alphapar}, using $\lambda=\sqrt{2/(1-\nu)}$ as the only fit parameter, giving a value for the Poisson ratio $\nu=0.5$, comparable to values expected for crystalline solids in two dimensions \cite{greaves2011poisson,mott2009limits}. It should be noted, however, that the profiles do not depend very sensitively on $\nu$, so that our determination of the Poisson ratio is not very precise.   As a final consistency check, we use the measured Youngs modulus and Poission ratio to predict the absolute values of the response functions, which now provide a reasonable quantitative match with the experimental results [Fig. \ref{fig:2} a-i]. This highlights that our theoretical description captures the main phenomena in a quantitative fashion. 

\begin{figure}[t!]
    \centering
    \includegraphics[width=\linewidth]{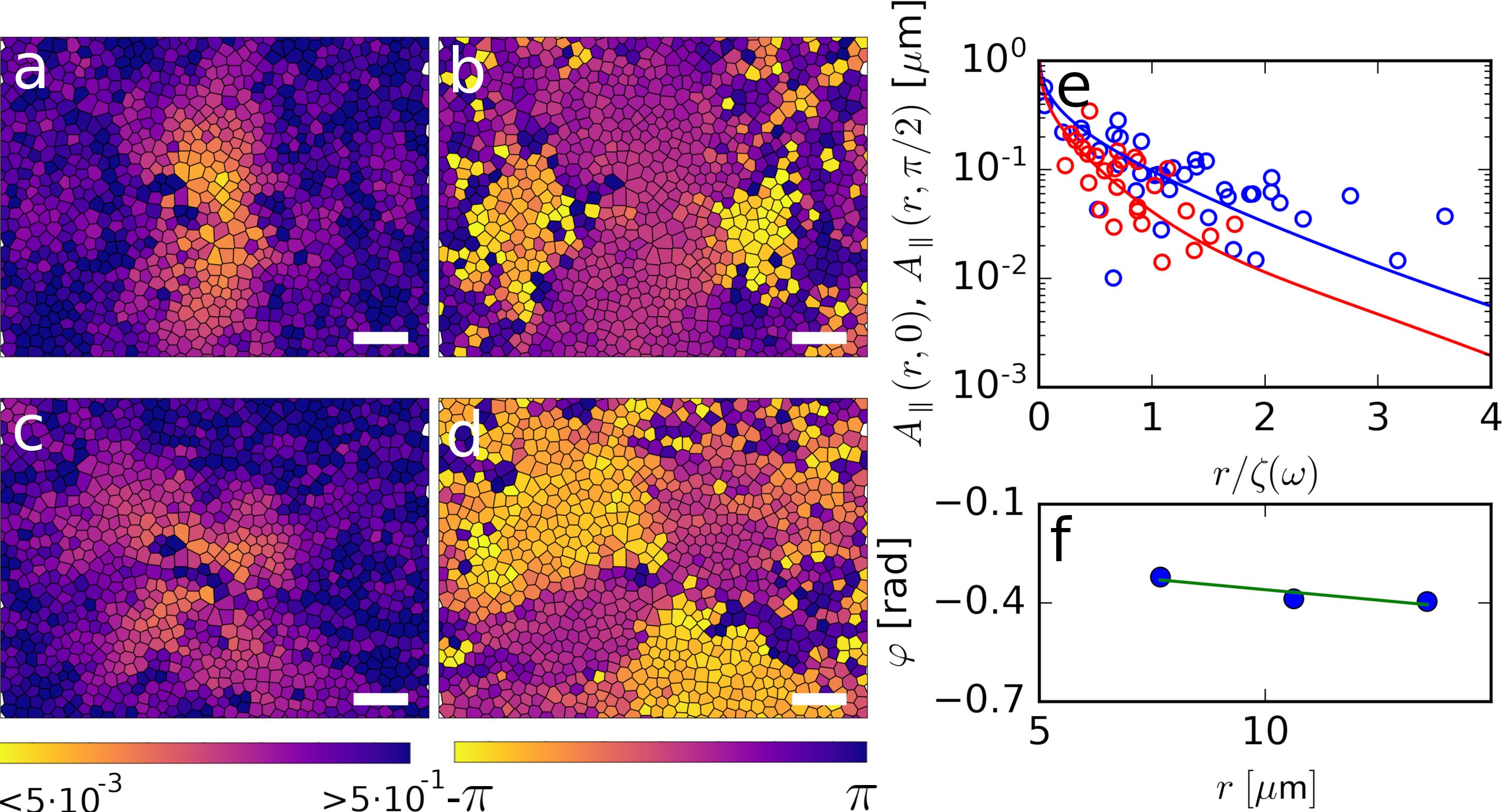}
    \caption{Experimental maps of (a) parallel displacement amplitude, (b) parallel displacement phase, (c) perpendicular displacement amplitude, and (d) perpendicular displacement phase of 2D colloidal glasses excited at $\omega=0.52$ rad/s. Amplitudes and phases have units micron and radians respectively, scalebars represent 20$\mu$m. (e) Superposition of parallel displacement amplitude of colloidal glasses for different frequencies along $\theta=0$ (blue) and $\theta=\pi/2$ (red) versus normalized distance, together with theoretical prediction. (g) Bin-averaged phase-distance plot to determine phase velocity for $\omega  =$  0.52 rad/s  in  the  parallel  direction.}
    \label{fig:4}
\end{figure}

Finally, to emphasize that our approach is not exclusive to ordered, crystalline, solids, we repeat the experiments described above for a disordered colloidal glass that lacks long-ranged order\cite{Vivek1850,hunter2012physics}. We prepare monolayers of a bi-disperse mixture (d=6.25 and 3.75 $\mu$m) of silica spheres by sedimentation. We confirm the absence of structural order from the liquid-like shape of the pair-correlation function and structure factor, while the particle dynamics are strongly arrested and caged as evidenced from their mean-squared displacement in the absence of external mechanical excitation (Supplementary Information 3). 

Despite the very different microstructure, we find that the displacement amplitude and phase patterns are very similar to those observed for the crystals (Fig.\ref{fig:4} a-d). The same analysis (Fig. \ref{fig:4}e,f) as for the crystals gives a Poisson ratio on the order of 0.5 and an elastic modulus on the order of $2\cdot10^{-7}$ N/m for the glasses. This is roughly a factor of three lower than for the crystals, which we attribute to non-affine softening due to the amorphous nature of glass\cite{PhysRevB.83.184205}. 

This highlights that the observation of mechanical waves at low $Re$ open up the way for mechanical characterization of both ordered and  disordered ultra-weak solids where conventional approaches fail. From Fig. \ref{fig:4}a-d it is also clear that the wave patterns are noisier for the glasses than for the crystals, which is probably due to their inhomogeneous structure, and therefore also inhomogeneous mechanical properties.  This clearly shows how elastic wave propagation is affected by the structure of the medium. We note that this is not yet captured by the theory, which assumes a homogeneous elasticity. 

Traditionally, microrheology is the method of choice to characterize the visco-elasticity of very weak elastic materials whose moduli are below the detection limit of conventional macroscopic rheometers. In microrheology, the visco-elastic features of the material are extracted from the motion of either actively-driven or thermally-excited (passive) tracer particles embedded in the  material \cite{PhysRevLett.74.1250, PhysRevLett.85.1774}. This analysis is based on the assumption that the generalized Stokes-Einstein relation holds, which  is only the case when the characteristic length scale of the material is much smaller than the probe particles. In particular for very inhomogeneous soft solids, this condition is often not met. As our approach is not bound by these limitations, the study of mechanical wave propagation at low $Re$ provides experimental access to the mechanical properties of extremely weak and strongly heterogeneous systems, such as marginal solids close to a mechanical critical point \cite{trappe2001jamming,broedersz2011criticality}.

In this paper we have shown how propagating elastic waves can be generated and detected at low $Re$ in ultrasoft solids, both ordered and disordered. Moreover, we have proposed an analytical model to describe and interpret the wave propagation, which is in excellent quantitative agreement with our experimental results. On the basis of this theory, a measurement of the wave's phase velocity and decay length gives access to the full linear elasticity of the material, giving values in excellent agreement with lattice theory predictions. In principle, our approach can be extended to probe the mechanics of ultrasoft three-dimensional materials such a biopolymer networks, where the Fourier-filtered elastic displacements can be obtained e.g. by embedding tracer particles in the material or by using digital image correlation approaches. This could open the way to characterize how localised mechanical signals acting on biological structures give rise to the complex spatio-temporal response that underlies mechanical communication in living organisms and on the role of local structures on the response of marginal networks. 

\section{Acknowledgements}
This work is part of the Industrial Partnership Program Hybrid Soft Materials that is carried out under an agreement between Unilever Research and Development B.V. and the Netherlands Organisation for Scientific Research (NWO). J.v.d.G. acknowledges the European Research Council for financial support (ERC Consolidator grant Softbreak). The work of J.S. is part of the VIDI research program (No. 723.016.001) of NWO.

\end{document}